\documentclass[10pt,twocolumn]{article}
\usepackage[top=50pt,bottom=70pt,left=40pt,right=40pt]{geometry}

\usepackage{hyperref}
\usepackage[table]{xcolor}
\usepackage{booktabs} 
\usepackage{amsmath,amssymb,amsfonts}
\usepackage{algorithmic}
\usepackage{graphicx}
\usepackage{xspace}
\usepackage{textcomp}
\usepackage{subcaption}
\usepackage{siunitx}
\usepackage{booktabs}
\usepackage{multirow}
\usepackage{subfiles}
\usepackage{xspace}

\usepackage{mathtools}
\usepackage{soul}

\graphicspath{{img/}{img/}}

\definecolor{boristext}{rgb}{0.22, 0.44, 0.88}
\definecolor{boriscomments}{rgb}{0.88, 0.04, 0.04}

\newcommand{\sys}{WACA\xspace}

\begin{document}

\title{Wi-Fi All-Channel Analyzer}

\author{Sergio Barrachina-Mu\~noz, Boris Bellalta, Edward Knightly}

\date{}

\maketitle

\begin{abstract}

In this paper, we present \sys, the first system to simultaneously measure the energy in all 24 Wi-Fi channels that allow channel bonding at 5 GHz with microsecond scale granularity.
With \sys, we perform a first-of-its-kind measurement campaign in areas including urban hotspots, residential neighborhoods, universities, and a sold-out stadium with 98,000 fans and 12,000 simultaneous Wi-Fi connections. The gathered dataset is a unique asset to find insights otherwise not possible in the context of multi-channel technologies like Wi-Fi. 
To show its potential, we compare the performance of contiguous and non-contiguous channel bonding using a trace-driven framework. We show that while non-contiguous outperforms contiguous channel bonding's throughput, occasionally bigger by a factor of 5, their average throughputs are similar.

    
\end{abstract}

\section{Introduction}
Channel bonding is a key mechanism for increasing Wi-Fi data rates as the maximum data rate increases in proportion to the total channel bandwidth. In Wi-Fi, while the \textit{basic} channel width remains 20~MHz, the maximum \emph{bonded} channel width has increased from 40~MHz in 802.11n \cite{11n} to 160~MHz in 802.11ac/ax \cite{11ac,tgax2019draft}, and 320~MHz in 802.11be \cite{11be}. During this time, the standard has evolved to not only support wider bandwidths, but also to enable more sophisticated channel bonding policies: in 802.11n,  only ``static''  channel bonding was allowed in which a fixed group of pre-configured channels must always be bonded.  Today, the standard enables a far richer set of capabilities including dynamic selection of channel width as well as bonding both contiguous and non-contiguous channels. 


In this paper, we make the following three contributions. First, we develop \sys as the first system to simultaneously measure all 24 Wi-Fi channels at 5 GHz with a 10~$\mu$sec sampling rate. \sys employs multiple synchronized WARP  software defined radios (SDRs) and has 24 antennas and 24 radio frequency (RF) chains to match the number of Wi-Fi channels that allow channel bonding (from channel 36 to 161) in the IEEE 802.11ac/ax standards. Our approach contrasts with prior work that uses one or several RF chains, thereby encountering \emph{second} scale channel switching delays. Thus, no prior methods capture all channels at the channel-access $\mu$sec  timescale. 

Second, using \sys, we conduct extensive measurement campaigns covering two continents, dense urban areas, and places of interest such as university campuses, apartment buildings, shopping malls, and hotels. We also perform measurements in the Futbol Club Barcelona's Camp Nou, one of the largest sports stadiums in the world. The shortest campaign took 20 minutes and the longest covered more than 1 week, and the total number of samples exceeds $10^{11}$. In all cases, we record signal strength on all channels at 10~$\mu$sec sample rate, from which we infer the epochs for which each channel is occupied.\footnote{The WACA dataset v1 can be found at \url{https://www.upf.edu/web/wnrg/wn-datasets}.}\footnote{All of the source code of WACA is open, encouraging sharing of algorithms between contributors and providing the ability for people to improve on the work of others under the GNU General Public License v3.0. The repository can be found at \url{https://github.com/sergiobarra/WACA_WiFiAnalyzer}.} Notice that the traces compose an unprecedented dataset to find insights otherwise not possible. We highlight Wi-Fi as the main research application of the dataset but others can also be taken into account, e.g., cognitive radios, coexistence between Wi-Fi and LTE, and coexistence between Wi-Fi and Internet of Things (IoT) technologies, to list a few examples.


Third, we introduce a trace-driven framework to evaluate the performance of channel bonding policies using the aforementioned high-resolution traces of channel activity. As the channel occupancies are highly dynamic, we group epochs according to their average utilization. While the stadium measurements have band occupancies as high as 99\%, even dense urban scenarios with many competing basic service sets (BSS's) yielded maximum occupancies near 45\%.

Finally, we use the trace-driven framework to compare the performance of contiguous vs. non-contiguous channel bonding and find that the increased flexibility of the latter can yield a throughput improvement of up to 5$\times$. Unfortunately, the scenarios for these gains occur quite rarely yielding modest average gains of less than 10\%, which may ultimately favor contiguous channel bonding, since it is simpler to implement.

Due to space constrains, we leave the study of key factors such as spectral correlation, bandwidth prevention, or standard compliant channel bonding policies as future work. 
\section{Related work}

\sloppy

Prior work performed spectrum measurements for Wi-Fi traffic, e.g., \cite{rademacher2018quantifying, taher2014global, subramaniam2015spectrum, wellens2009empirical, kone2012effectiveness, day2014activity, salahdine2017real, biswas2015large, 6030801, kone2010feasibility, hoyhtya2016spectrum}. Example objectives include creating interference maps \cite{hoyhtya2016spectrum}, assessing interference behavior \cite{hanna2012distributed}, surveying Wi-Fi usage  \cite{day2014activity}, quantifying spectrum occupancy in outdoor testbeds \cite{rademacher2018quantifying,taher2014global}, designing efficient scanning methods \cite{subramaniam2015spectrum, salahdine2017real}, modeling spectrum use  \cite{wellens2009empirical},  opportunistic spectrum access \cite{kone2012effectiveness, kone2010feasibility}, dynamic channel selection \cite{hanna20143}, and assessing real-world network behavior by examining data from thousands APs \cite{biswas2015large}.
Unfortunately, no such prior work provided simultaneous measurement across the entire 5 GHz band, which we require for our channel bonding study. While some papers do provide multi-channel measurements, e.g., \cite{yin2012mining,kone2010feasibility, kone2012effectiveness,rademacher2018quantifying}, they do so via sequential scanning, thus taking on the order of seconds to change from one channel to the next, orders of magnitude beyond the transmission time scale for channel bonding. Namely, \sys measures all channels simultaneously using SDRs having a sampling rate of $10 \mu$sec. Moreover, prior work does not consider (for example) stadiums.

Throughput gains of channel bonding have been demonstrated previously in testbeds. In particular, IEEE 802.11n static channel bonding has been shown to be affected not only by link signal quality, but also by the power and rates of neighboring links \cite{deek2011impact}. Likewise, intelligent channel bonding management was shown to benefit from identifying the signal strength of neighboring links and interference patterns \cite{deek2014intelligent}. High bandwidths were shown to be vulnerable to increased thermal noise or the power per Hertz reduction \cite{arslan2010auto,arslan2013acorn}. Nonetheless, existing experimental results have targeted only on one or few controlled links at most. In contrast, we develop \sys as a monitoring system in order to measure multiple BSS's in various operational settings, which allows tackling how a channel bonding BSS interacts with surrounding BSS's in a broad set of scenarios.

Simulation studies and analytical models have also been employed to study channel bonding, e.g., early simulation studies demonstrated throughput gains of channel bonding compared to single-channel transmission \cite{park2011ieee, gong2011channel}. Likewise, analytical models have been proposed to study channel bonding, especially through Markov chains \cite{bellalta2015interactions,yazid2019stochastic}. Analytical models for unsaturated traffic have been also proposed \cite{kim2017throughput, barrachina2019overlap}. As well, different channel bonding policies were introduced and modeled for spatially distributed scenarios \cite{barrachina2019dynamic}. However, such work does not have an experimental validation as presented here. In this regard, \sys datasets can be used for validation of the aforementioned methods under real-world channel occupancies in future work.

\section{Wi-Fi All-Channel Analyzer}   
\label{sec:acswa}

	\begin{figure*}[t!!!]
		
		\centering
		\begin{subfigure}{0.68\textwidth}
			\includegraphics[width=\textwidth]{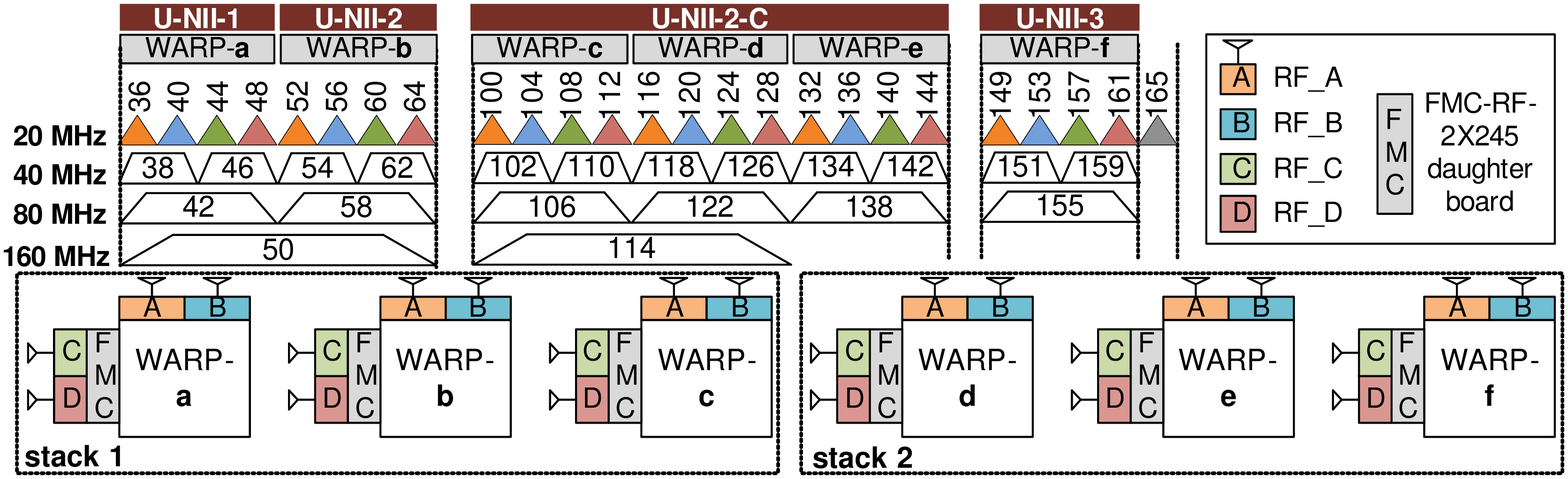}
			\caption{IEEE 802.11ac/ax channelization at the 5-GHz and assignation per RF.}
			\label{fig:warp_scheme}
		\end{subfigure}
		\hspace{3mm}
		\begin{subfigure}{0.277\textwidth}
			\includegraphics[width=\textwidth]{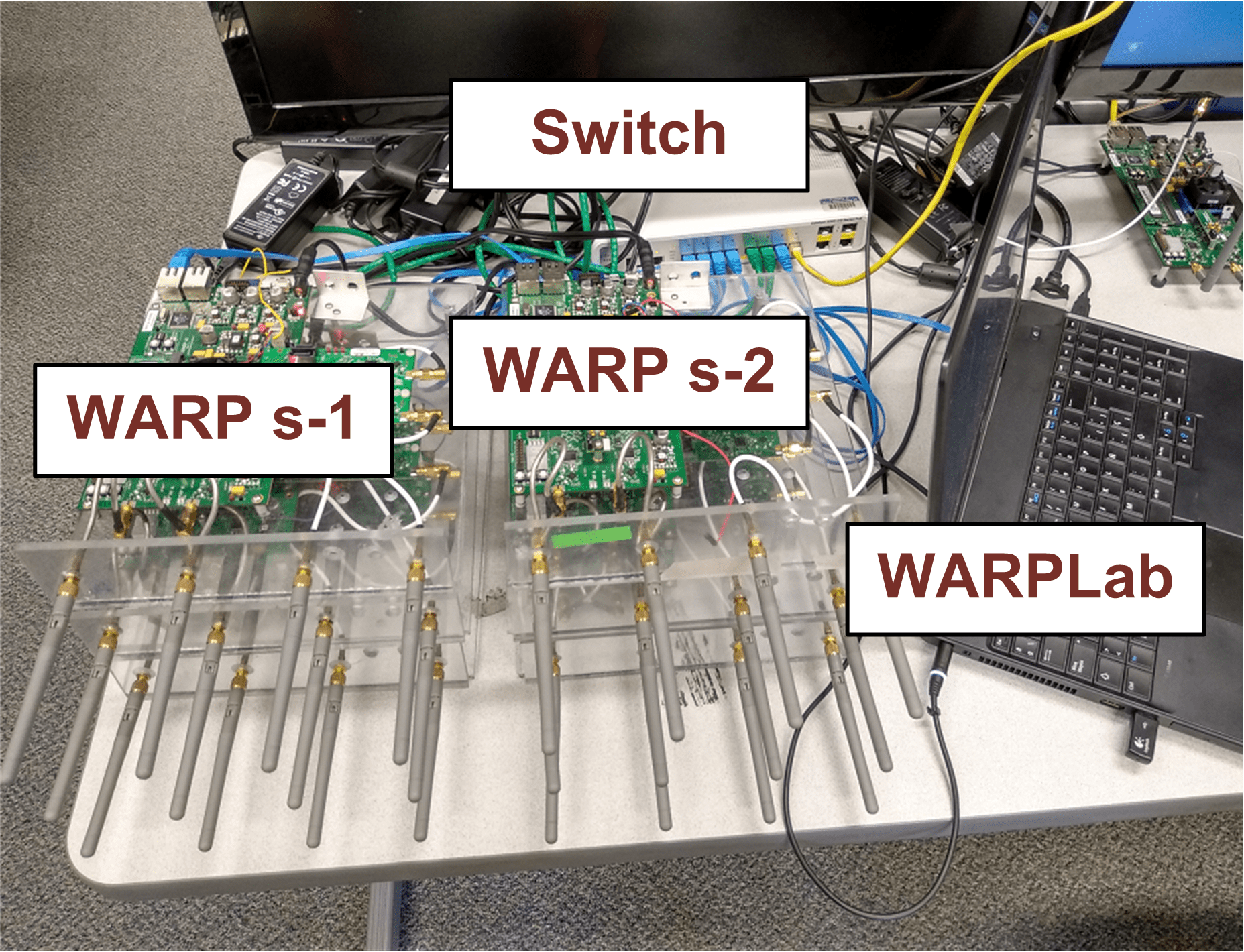}
			\caption{Deployment schematic.}
			\label{fig:warp_deployment}
		\end{subfigure}
		
		\caption{The \sys all WiFi channels spectrum analyzer.}
		\label{fig:warp_scheme_deployment}
	\end{figure*}

\subsection{Overview}
	
Our objective is to simultaneously capture activity on all Wi-Fi channels, i.e., all 24 basic (non aggregated) 20 MHz channels in the 5 GHz band that permit channel bonding. In principle, this could be achieved with an  off-the-shelf spectrum analyzer. However, most spectrum analyzers are not capable of dealing with the required bandwidth of this objective, i.e., they cannot  simultaneously measure the entire Wi-Fi 5-GHz band: 645 MHz ranging from channel 36 to 161 (i.e., from 5170 to 5815 MHz). Moreover, wide-band spectrum analyzers that can cover this bandwidth lack resolution to analyze basic channels within the band. 

Likewise, one could envision a system comprising 24 off-the shelf Wi-Fi cards as sniffers, one per basic channel. Unfortunately, such a system would be quite unwieldy and would introduce a challenge of ensuring synchronicity among wireless cards: restricting the delay between channel measurements to the order of nano/microseconds is unfeasible due to the hardware interrupt latency and jitter from the different peripherals \cite{buttazzo2016design,farzinvash2009scheduling}.	

Thus, we design \sys to simultaneously measure power (and I/Q signals if required) on all 5-GHz basic channels. Key benefits of \sys include the simplicity of experimental procedures (from deployment to post-processing), a dedicated RF chain per channel (covering the whole band and easing hardware failure detection), and the ease of adaptation/configuration empowered by the WARPLAb framework~ \cite{anand2010warplab}.
	
\subsection{Building Blocks}
	
	
The key building blocks of \sys are \textit{i}) six WARP v3 programmable wireless SDRs  \cite{warpProject}, \textit{ii}) six FMC-RF-2X245 dual-radio FMC daughterboards,\footnote{FMC-RF-2X245 datasheet: \url{https://mangocomm.com/products/modules/fmc-rf-2x245}, retrieved January 30, 2020.} \textit{iii}) 24 5-GHz antennas (one per RF chain), and \textit{iv}) one Ethernet switch to enable communication between the WARPLab host (e.g., PC) and the WARP boards. The preeminent building block is WARP, a scalable and extensible programmable wireless platform to prototype advanced wireless networks. The FMC-RF-2X245 module dual-radio FPGA Mezzanine Card (FMC) daughterboard extends the capability of WARP v3 boards from 2 to 4 RF chains. Therefore, by combining 2 stacks of 3 WARP boards each with their corresponding FMC-RF-2X245 daughterboards, we realize 24 RF chains (with one 5-GHz antenna each) enabling us to assign a single RF chain per basic channel. Finally, the Ethernet switch enables the communication from the WARP nodes to the WARPLab host. Figure~\ref{fig:warp_scheme} shows the assignment of the RF chains to each basic channel allowed for bonding and Figure~\ref{fig:warp_deployment} depicts the physical realization of \sys.
	
\subsection{Measurement Methodology}
An iterative procedure is followed for collecting power samples. Namely, in each iteration,  \sys first simultaneously measures the power at each basic channel during $T_\text{m}$ and then takes $T_\text{proc}$ to process and forward the measurements to the WARPLab host. Both tasks are sequentially performed until the end of the measurement campaign.
Table \ref{table:warplab_setup_params} shows the main WARPLab parameters used for measurements. Notice that all parameters are fixed except $N_\text{it}$, used for determining the duration of the measurement campaign.

	\begin{table}[t!]
		\begin{tabular}{@{}cc@{}}
			\toprule
			\textbf{Parameter}  & \textbf{Value}                                                                                             \\ \midrule
			Active RFs (channels 36 to 161), $\mathcal{R}$  & $\{1,2,...,24\}$                              
			\\
			Iteration's measurement duration, $T_\text{m}$  & 1 s   
			\\
			Iteration's processing duration, $T_\text{proc}$ & $\sim 9$ s
			\\
			Duration of a complete iteration, $T_\text{it}$ & $\sim 10$ s
		    \\
			\begin{tabular}{@{}c@{}}Original no. of power samples \\ per channel per second, $n_\text{s}^*$\end{tabular} & $10^7$
			\\
			\begin{tabular}{@{}c@{}}Downsampled no. of power samples \\ per channel per second, $n_\text{s}$\end{tabular} & $10^5$
			\\
			Time between consecutive samples, $T_\text{s}$ & 10 $\si\micro$s
		    \\
			\midrule
			Number of iterations, $N_\text{it}$ & 125-59500  \\
			\bottomrule
		\end{tabular}
		\caption{WARPLab setup.}
		\label{table:warplab_setup_params}
	\end{table}
	
	WARP boards install the MAX2829 transceiver, which has a fixed 10 Msps received signal strength  sampling rate. Accordingly, since the measurement duration in an iteration is $T_\text{m}$ = 1 second, the number of consecutive samples captured per basic channel per iteration is $n_\text{s}^* = 10^7$. Then, in each iteration, we store \mbox{$|\mathcal{R}|\times n_\text{s}^* = 24\times10^7$} samples.
    Nonetheless, to decrease the amount of required memory, we downsample the gathered samples in each iteration by a 100$\times$ factor, thus reducing the data size per iteration from 60 MB to 600 KB. Essentially, while the transceiver measures 1 power sample every 100 ns by default, we keep just 1 sample every $T_s = 10$~$\si\micro$s. Notice that the resulting time scale is also suitable given Wi-Fi timings. Indeed, the short interframe space (SIFS) is the smallest interframe space and takes 16~$\si\micro$s~($>T_s$).
	
	
	As for the duration of processing and forwarding (period in which no data is collected), $T_\text{proc}$ entails a significant yet unavoidable delay overhead with respect to the total duration of an iteration $T_\text{it} = T_\text{m} + T_\text{proc}$. Specifically, for the host PC used in all the scenarios in the dataset (Intel Core i5-4300U CPU @ 1.9 GHz x 4 and 7.7 GiB memory) and the parameters listed in Table \ref{table:warplab_setup_params}, $T_\text{proc} \approx 9$ s for $T_\text{m} = 1$~s. 
	Once initiated, \sys operates by itself, and no human intervention is required.
	
	
	
\subsection{Validation} \label{sec:validation}

\begin{figure}[ht]
	\centering
	\includegraphics[width=0.4\textwidth]{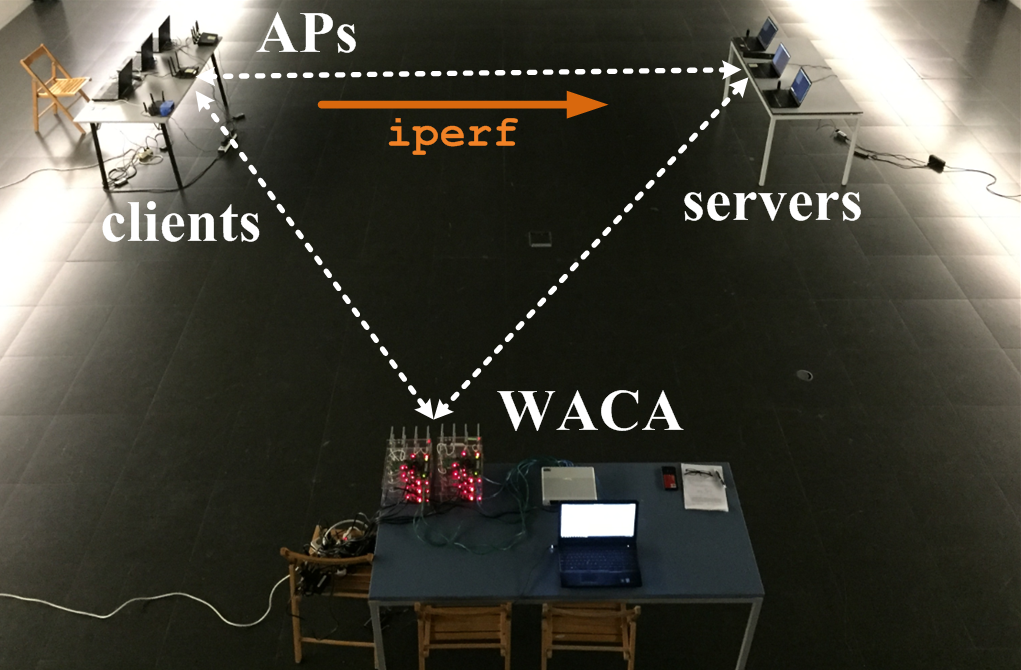}
	\caption{Validation testbed at the main events room of Universitat Pompeu Fabra's Communication campus.}
	\label{fig:testbed}
\end{figure}

Before deploying \sys for the measurement campaigns, we perform an extensive set of in-lab controlled experiments for validation, comprising over $6\times10^8$ power samples collected in \sys cross validated with controlled and known transmissions from commercial APs. We conduct the validation of all the boards by first measuring the power perceived against distance and transmission bandwidth. Then, we explore the spectrum behavior when setting up different off-the-shelf channel bonding configurations. The measurements were gathered in an empty and large event room (about 300 m$^2$). Figure \ref{fig:testbed} shows the complete deployment. In particular, we deploy \sys including the PC hosting the WARPLab framework (lower part of the figure), 3 PCs receiving \texttt{iperf} traffic (right part of the figure), and 3 PCs sending \texttt{iperf} traffic with the corresponding 3 APs enabling the \texttt{iperf} connections (left part of the figure). The AP models we use are Asus RT-AC87U (AS), TP-Link AC1750 (TP), and Linksys WRT 3200 ACM (LS). Remarkably,  all of these APs are only capable of performing static channel bonding.

\subsubsection{RSSI vs. central frequency}

Ideally, we would like each of the 24 RFs to perceive exactly the same RSSI value when receiving the same signal modulated at a central frequency $f$. Besides, we would also like each of the RFs to perceive exactly the same RSSI value at every central frequency $f$ when the same signal is modulated at each $f$. In other words, the RSSI vs. $f$ should be the same flat curve for all the RFs. In order to validate such premises, we use a WARP node acting as a transmitter and sequentially connect the transmitter RF (which keeps generating the same 20-MHz signal) to each of the 24 RFs of WACA for every channel from 36 to 161 via an RF-SMA connector. As shown in Figure 3, a similar RSSI is perceived at every RF. Besides, a similar flat frequency response is achieved for every RF.

\subsubsection{RSSI vs. distance}

\begin{figure}[t!]
	\centering
	\begin{subfigure}{0.44\textwidth}
		\centering
		\includegraphics[width=\textwidth]{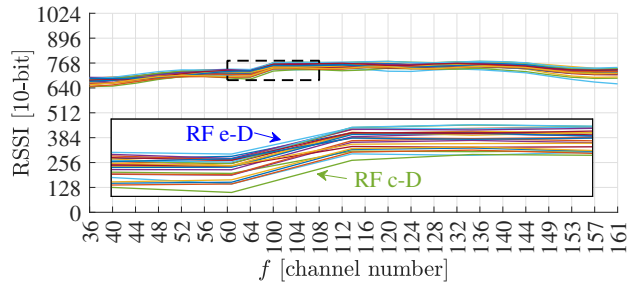}
		\caption{RSSI vs. central frequency. Each curve represents an RF.}
		\label{fig:rssi_central_frequency}
	\end{subfigure}
	\vskip 1 mm
	\begin{subfigure}{0.44\textwidth}
		\centering
		\includegraphics[width=\textwidth]{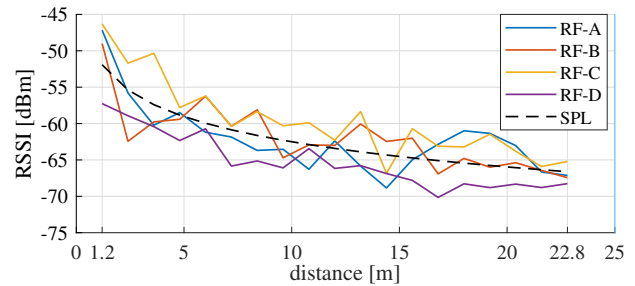}
		\caption{RSSI vs. distance. The curve named SPL stands for the simplified path loss model with $\alpha=1.15$.}
		\label{fig:validation_distance}
	\end{subfigure}
	\vskip 1 mm
	\begin{subfigure}{0.44\textwidth}
		\centering
		\includegraphics[width=\textwidth]{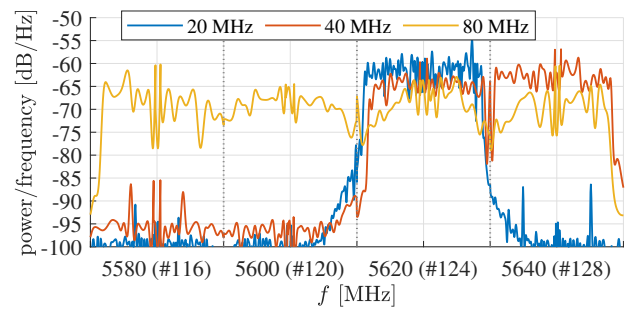}
		\caption{Transmission bandwidth effect on power.}
		\label{fig:validation_bandwidth}
	\end{subfigure}
	
	\caption{Validation metrics.}
	\label{fig:val_metrics}
	
\end{figure}

We measure the RSSI in dBm\footnote{We follow the MAX2829 transceiver data sheet to convert from 10-bit RSSI values to dBm values as done in \cite{jang2018post}.} at each basic channel at different distances $d$ from the AP to \sys  ranging from 1.2 to 22.8 meters. We took $10^5$ samples per point. For redundancy, we use 4 RFs to measure the same channel $p=124$.\footnote{We selected this channel since negligible activity was detected.} As expected, the RSSI tends to decrease with the distance due to the path loss effect as shown in Figure~\ref{fig:validation_distance}. For the sake of representation, we also plot the simplified path loss (SPL) of the mean power received at each point: $\text{RSSI}(d) = \text{RSSI}(d_0)-\alpha 10\log_{10}(d)$, with $d = 1$ m. Despite the experiments were conducted in a large, empty room without furniture, we can see the multipath effects in the sparse RSSI values perceived by each RF due to the reflection of the walls. Such effect is anticipated since the minimum separation between RFs is 6 cm.

\subsubsection{RSSI vs. bandwidth}


This experiment aims at measuring the RSSI perceived at a fixed distance (3.6 m) in four basic channels at the same time for 20, 40, and 80 MHz transmissions. The general procedure followed in the experiment can be summarized in 4 steps: \textit{i)} the AS AP is set up with bandwidth $b \in \{20, 40, 80\}$ MHz and primary channel $p=124$; \textit{ii)} an \texttt{iperf} client is placed close to the AP and associates to it; \textit{iii}) an \textit{iperf} server is located at distance $d = 3.6$ m from the AP and also associates to it; finally, \textit{iv}) the \texttt{iperf} communication is triggered and \sys captures the resulting RSSI at channels 116, 120, 124, and 128, thus covering 80 MHz. Notice that it is enough to use just one WARP board with the corresponding daughterboard to cover the 4 basic channels of interest. 
Theoretically, when doubling the bandwidth, the transmission power gets reduced by a half (3 dB). We corroborate this fact by looking at the similar reduction factor in the power detected per Hz in Figure~\ref{fig:validation_bandwidth}.

\subsubsection{Controlled testbed evaluation} \label{sec:testbed}

We now measure the activity of 3 overlapping BSSs' with static channel bonding capabilities under different traffic regimes. As shown in Figure~\ref{fig:testbed}, we separated the APs from their corresponding \texttt{iperf} servers by the same distance $d=4.8$ m. The \sys platform was placed equidistantly from the central AP and STA also at $d=4.8$ m.


We analyze two particular setups where none and all the BSSs' saturate. The load of every BSS is 20 and 150 Mbps in the first and second setup, respectively. Table~\ref{table:controlled_experiments} collects the setups' details. Essentially, in the unsaturated scenario, all the BSSs' share the same primary channel (48) but are set with different bonding capabilities. The second setup assigns different primary channels and bandwidth capabilities to each BSS. Note that we changed the primary channels from the previous experiments due to AP hardware restrictions. Nonetheless, we also confirmed that the external interference at the new band of interest was negligible.

\begin{table}[t!]
	\centering
	\small
	\begin{tabular}{@{}ccccccc@{}}
		\toprule
		Regime                                                                                   & AP & 36                                 & 40                                 & 44                       & 48                                 & Thr. [Mbps] \\ \midrule
		& AS & \cellcolor{green}           & \cellcolor{green}           & \cellcolor{green} & \cellcolor{green}\textbf{p} & 20                       \\
		& TP &                                    &                                    & \cellcolor{red} & \cellcolor{red}\textbf{p} & 20                     \\
		\multirow{-3}{*}{\begin{tabular}[c]{@{}c@{}}Unsaturated\\ $\ell = 20$ Mbps\end{tabular}} & LS &                                    &                                    &                          & \cellcolor{yellow}\textbf{p} & 20                     \\
		& AS & \cellcolor{green}           & \cellcolor{green}\textbf{p} &                          &                                    & 100                      \\
		& TP & \cellcolor{red}\textbf{p} &                                    &                          &                                    & 59                       \\
		\multirow{-3}{*}{\begin{tabular}[c]{@{}c@{}}Saturated\\ $\ell = 150$ Mbps\end{tabular}}    & LS & \cellcolor{yellow}           & \cellcolor{yellow}           & \cellcolor{yellow} & \cellcolor{yellow}\textbf{p} & 16                     \\ \bottomrule
	\end{tabular}
	\caption{Setup of controlled experiments. Letter p indicates the primary channel whereas colors represent the allocated bandwidth to each AP.}
	\label{table:controlled_experiments}
\end{table} 

\begin{figure}[t!]
	\centering
	\begin{subfigure}{0.23\textwidth}
		\centering
		\includegraphics[width=\textwidth]{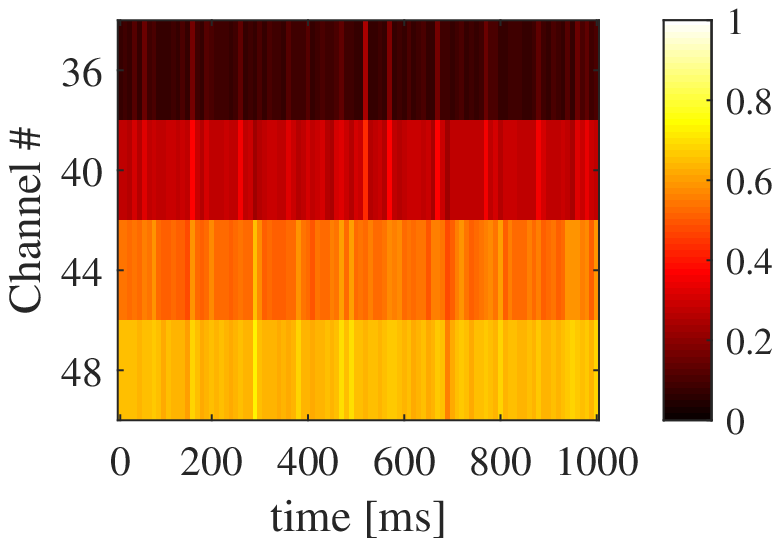}
		\caption{Unsaturated.}
		\label{fig:spectogram_controlled_unsat}
	\end{subfigure}
	~
	\begin{subfigure}{0.23\textwidth}
		\centering
		\includegraphics[width=\textwidth]{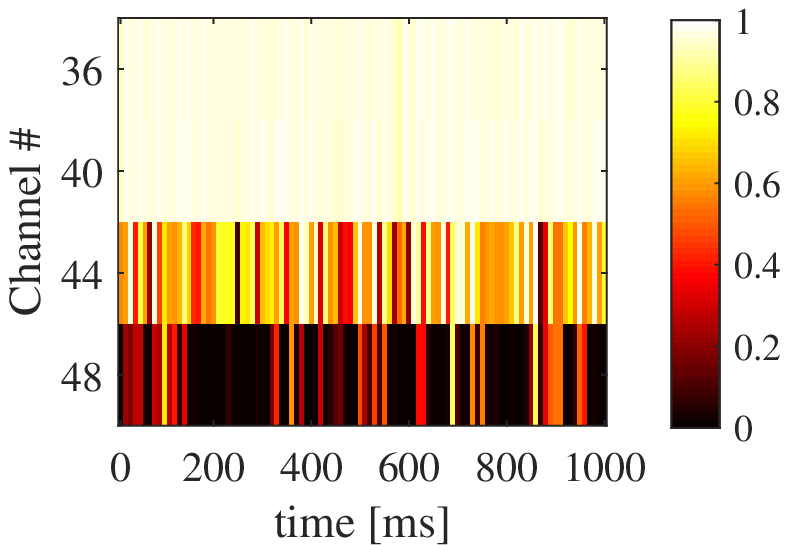}
		\caption{Saturated.}
		\label{fig:spectogram_controlled_sat}
	\end{subfigure}
	
	\caption{Spectogram evolution of the controlled scenarios. Slots represent the 10-ms mean occupancy.}
	\label{fig:spectograms}
	
\end{figure}

\sys collected 1 second of consecutive RSSI samples during the execution of each setup to measure the spectrum occupancy. 
%
Figure~\ref{fig:spectograms} shows the occupancy evolution at each basic channel for the unsaturated and saturated setups. We confirm that the former setup does not reach full occupancy in any channel, whereas it is the contrary for the later. Indeed, we observe that the load \mbox{($\ell=150$~Mbps)} is high enough to make channels 36 and 40 occupied almost all the time (i.e., occupancy at the first 40-MHz band is always close to 1) while not being able to successfully deliver all the load as indicated by the throughput values in Table~\ref{table:controlled_experiments}. This has an important consequence for the LS AP. Essentially, since static channel bonding is applied, the LS AP does not benefit from having its primary channel (48) far from the rest of APs. Consequently, its throughput performance is drastically reduced given that AS and TP introduce inadmissible activity in the first 40 MHz band.

\section{Measurement Campaigns} 
\label{sec:dataset}

Using \sys, we perform extensive measurement campaigns covering two continents, dense urban areas, and multiple hours of samples in places of interest such as university campuses, apartment buildings, shopping malls, hotels and the Futbol Club Barcelona (a.k.a Bar\c{c}a) stadium (Camp Nou), one of the largest sports stadiums in the world. The measurements were taken from February to August 2019 in Houston, TX, USA, and Barcelona, Spain. The shortest experiment took 20 minutes and the longest covered more than 1 week. The list of locations is shown in Table~\ref{table:locations}.

\begin{table}[h!]
	\small
	\begin{tabular}{@{}cccc@{}}
		\toprule
		\textbf{Id} & \textbf{Location}                                                         & \textbf{Type} & \textbf{Duration} \\ \midrule
		\texttt{1\_RVA}      & Rice Village Apart., HOU                                              & Apartment     & 1 day             \\
		\texttt{2\_RNG}     & RNG lab at Rice, HOU                                                      & Campus        & 1 day             \\
		\texttt{3\_TFA}      & Technology for All, HOU                                                   & Com. center   & 1 day             \\
		\texttt{4\_FLO}      & \begin{tabular}[c]{@{}c@{}}Flo Paris, HOU\end{tabular} & Cafe          & 1 hour            \\
		\texttt{5\_VIL}      & \begin{tabular}[c]{@{}c@{}}Rice Village, HOU\end{tabular}  & Shopping mall & 20 min            \\
		\midrule
		\texttt{6\_FEL}      & La Sagrera, BCN                                              & Apartment     & 1 week            \\
		\texttt{7\_WNO}      & WN group office, BCN                                                & Campus        & 1 day             \\
		\texttt{8\_22@}      & 22@ area, BCN                                                                          & Office area              & 1 day                  \\
		\texttt{9\_GAL}       & Hotel Gallery, BCN                                                                          & Hotel             & 1 day        
		\\
		\texttt{10\_SAG}       & Sagrada Familia, BCN                                                                          & Apartment             & 4 days          
		\\
		\texttt{11\_FCB}      & Camp Nou, BCN                                                                          & Stadium              & 5 hours                  \\ \bottomrule
	\end{tabular}
	\caption{Measurement locations.}
	\label{table:locations}
\end{table}

\begin{figure}[t!]
	\begin{subfigure}{.475\textwidth}
		\centering
		\includegraphics[width=1\textwidth]{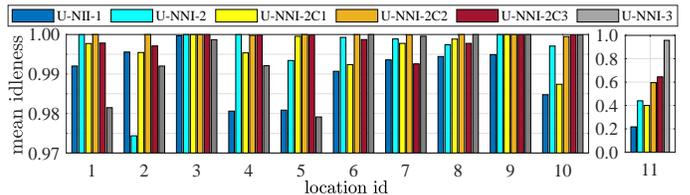}
		\caption{Mean idleness per band. }
		\label{fig:mean_idleness}
	\end{subfigure}
	\begin{subfigure}{.475\textwidth}
		\centering
		\includegraphics[width=1\textwidth]{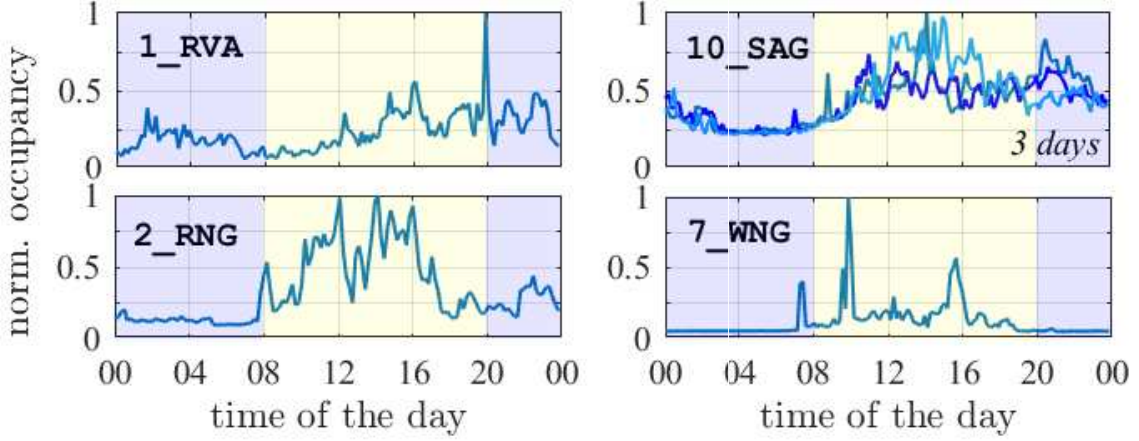}
		\caption{Temporal evolution of the normalized occupation for apartments (upper) and universities (lower) highlighting day (yellow) and night (blue).}
		\label{fig:results_temporal_occupancy}
	\end{subfigure}
	\caption{General spectrum occupancy trends.}
	\label{fig:general_plots}
\end{figure}
Here, we overview the complete dataset: there are 153,033 iterations of 1 second accounting for 42 hours, 30 minutes, and 3 seconds of actual measurements in the 24 20-MHz channels of the 5 GHz WiFi band.
We first assess the entire activity record of each location at band $\mathcal{B}$, where $\mathcal{B}$ is a predefined set of contiguous channels. For instance, U-NII-1 band is defined as $\mathcal{B} = \{1,2,3,4\}$, corresponding to basic channels 36, 40, 44, and 48 (see Figure \ref{fig:warp_scheme}).\footnote{The rest of bands are sequentially composed of groups of four consecutive 20-MHz channels. So, the next band is U-NNI-2 with channels $\{5,6,7,8\}$ (basic channels 52, 56, 60, and 64), and the last one is U-NNI-3 with channels $\{21,22,23,24\}$ (basic channels 149, 153, 157, 161).} Figure~\ref{fig:mean_idleness} shows the normalized mean idleness of each band, i.e., the mean number of samples that were found idle in each channel of band $\mathcal{B}$. We observe that the spectrum is idle most of the time in all scenarios except the stadium, indicating that the 5 GHz band is still profoundly underutilized even in densely populated areas.

Figure~\ref{fig:results_temporal_occupancy} shows the daily temporal evolution of 4 example locations (2 apartments in the upper subplots, and 2 university campuses in the lower subplots). For the sake of representation, we plot the normalized occupancy of the whole band averaged in periods of 10 minutes. Concretely, we normalize with respect to the highest 10-minutes average occupancy encountered in each location. We clearly observe higher activity at working hours in the campus locations and a much less variable pattern in the apartment locations. In any case, from the low  spectrum utilization observed in Figure~\ref{fig:mean_idleness}, significant opportunities for channel bonding can be expected.

\begin{figure}[ht!]
		\centering
		
		\begin{subfigure}{0.22\textwidth}
			\centering
        	\includegraphics[width=\textwidth]{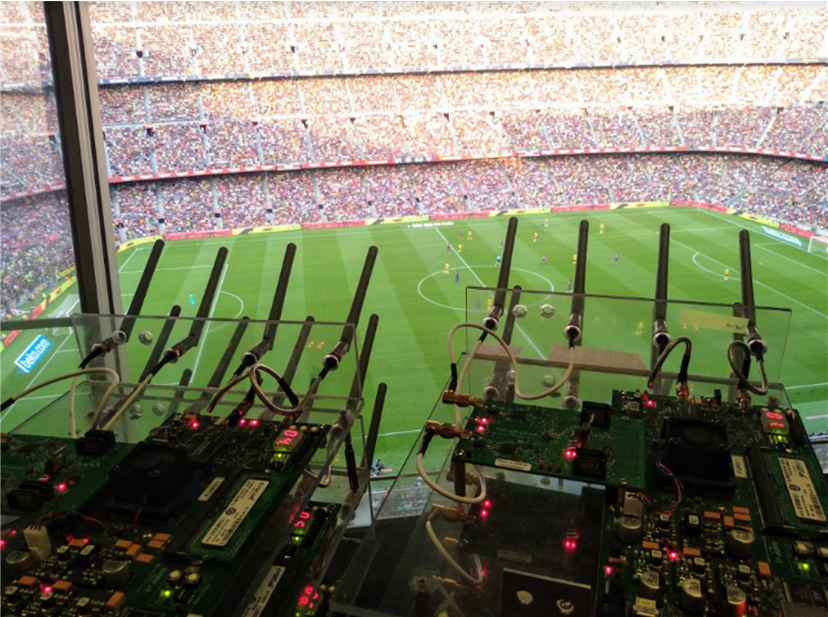}
    		\caption{\sys deployment.}
    		\label{fig:fcb_location}
		\end{subfigure}
		~
		\begin{subfigure}{0.225\textwidth}
			\centering
			\includegraphics[width=\textwidth]{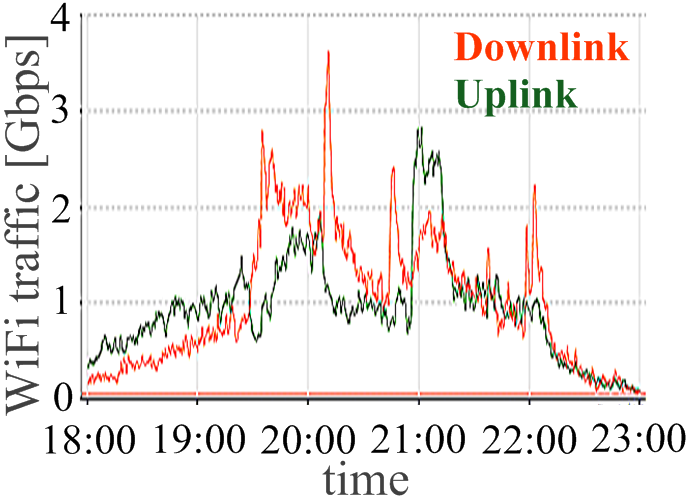}
			\caption{Traffic evolution.}
			\label{fig:fcb_traffic}
		\end{subfigure}
		
		\begin{subfigure}{0.46\textwidth}
			\centering
			\includegraphics[width=1\textwidth]{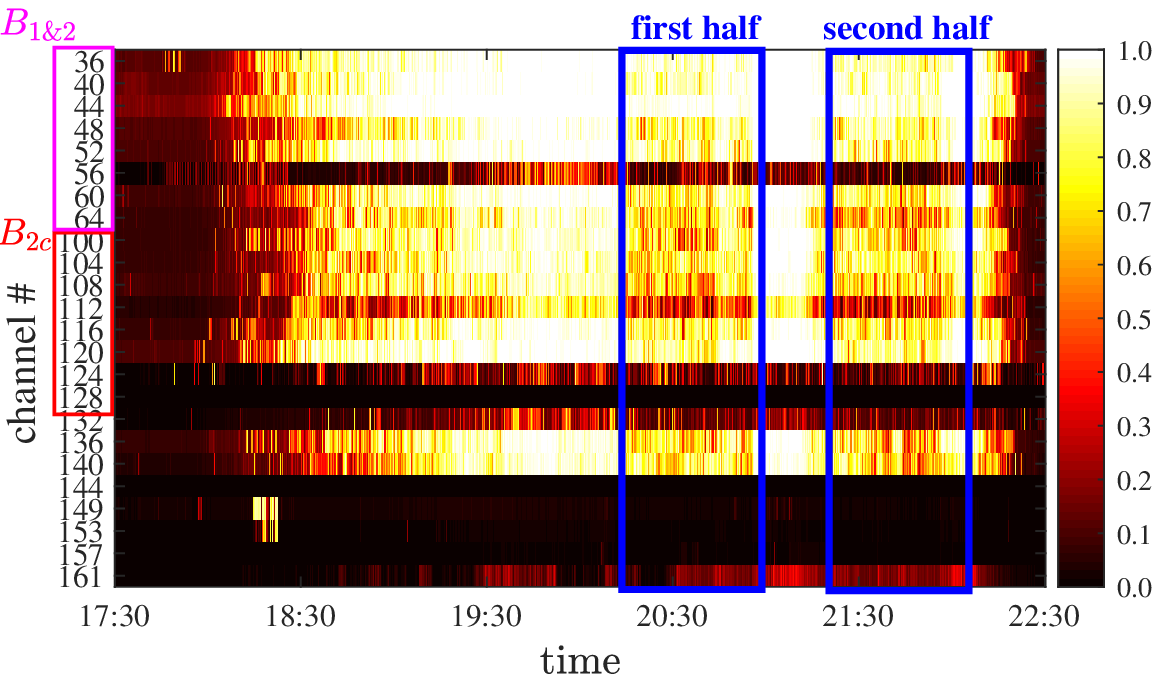}
			\caption{Spectrum occupancy evolution.}
			\label{fig:fcb_occupancy_temporal}
		\end{subfigure}
		
		\caption{Wi-Fi activity at the Camp Nou stadium: a) \sys deployment in the press box, b) downlink and uplink Wi-Fi traffic evolution (source: stadium's IT management), and c) spectrum occupancy evolution.}
		\label{fig:fcb_activity_plots}
	\end{figure}

Figure~\ref{fig:fcb_location} shows a photograph of the WACA setup in the Futbol Club Barcelona's stadium. We deployed \sys in the press box of the stadium during a football game with over 98,000 spectators present. Measurements were taken on August 4, 2019, from 17:24 to 22:30 accounting for a total duration of 5~hours and 6 minutes. On that date, the Joan Gamper trophy was held, which pitted the local club (Futbol Club Barcelona)
against the visiting club (Arsenal Football Club). Free Wi-Fi was provided to the audience through thousands of APs primarily installed beneath the seats. We also obtained data from the stadium's network management team which indicated that up to 12,000 Wi-Fi clients were simultaneously connected. Downlink and uplink traffic is depicted in Figure~\ref{fig:fcb_traffic}\footnote{Aggregated downlink and uplink traffic was provided by Futbol Club Barcelona's IT management.} and the spectogram in Figure~\ref{fig:fcb_occupancy_temporal}. Each slot in the spectogram represents the occupancy of each channel averaged in 1-second periods. We observe that most  channels were highly occupied during the measurements. Moreover, we can observe the users' behavior induced nonstationarity of the traces as the match progressed. Namely, while activity is always high, there is a notable reduction during  play (first and second half) compared to activity before, between halfs, and right after the game time.
We also observe that the majority of the channels are crowded most of the time with periods reaching mean band occupancy values rising up to 99\%. In fact, 22\% of the periods are above 80\% occupancy.
	
\section{Evaluation of bonding gains}

In this section, we study the performance of channel bonding against the band occupancy. Notice that, due to the disparate nature of the stadium scenario (recall the contrasting mean idleness in Figure~\ref{fig:mean_idleness}), we focus our study on the non-stadium measurement campaigns to provide insights on a more homogeneous dataset. The analysis of the stadium campaign will be covered in future work.

\subsection{Evaluation Methodology}

\subsubsection{Overview}
\begin{figure}[b!]
	\centering
	\includegraphics[width=0.38\textwidth]{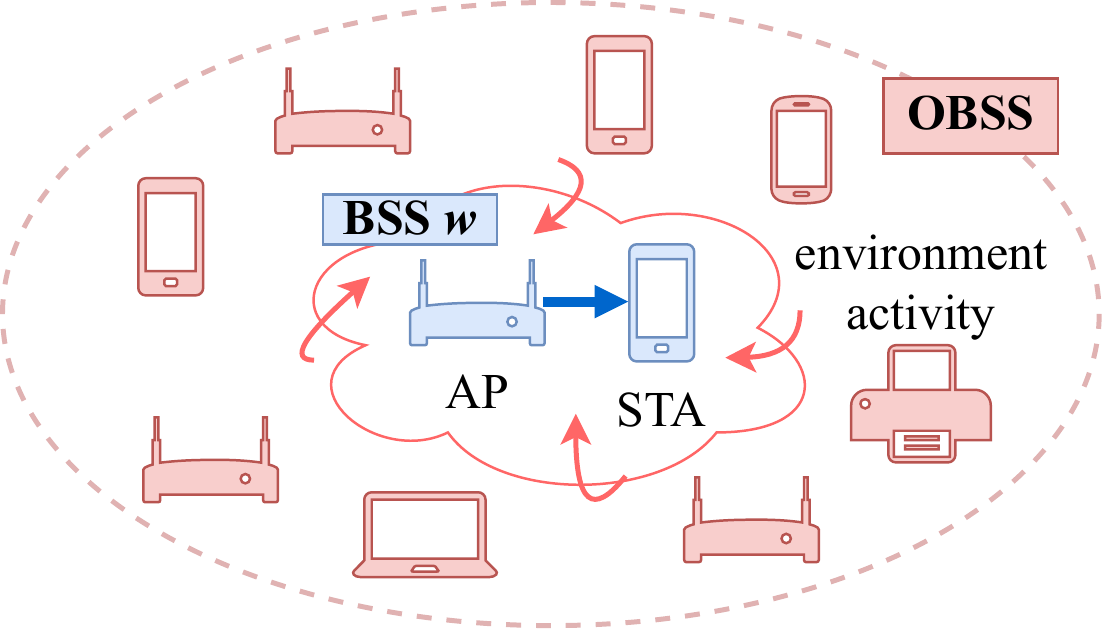}
	\caption{Diagram of the system model. The fully backlogged channel-bonding BSS, $w$, operates under instances of the same environment -- or overlapping BSS (OBSS) -- that WACA perceived at the measurement campaigns.}
	\label{fig:system_model}
\end{figure}

Our objective is to study the throughput performance that a fully backlogged channel-bonding BSS, $w$, would obtain if it encountered the channels recorded in the measurement campaigns described above. An schematic of the system model is illustrated in Figure~\ref{fig:system_model}. BSS $w$ consists of an access point (AP) and a one or multiple stations (STAs) sufficiently close to assume that they perceive the same RSSI measurements.

The throughput is a function of a number of factors such as which primary channel the bonding BSS selects, which channel bonding policy it employs, and the spectrum occupancy perceived while operating. So, here we describe the trace-driven framework we use. The RSSI measurements captured by \sys at scenario $s$ are represented by a 2-dimensional matrix $Y^s$ of size $(N_\text{it}^s \times n_\text{s})\times |\mathcal{R}|$, where $N_\text{it}^s$ is the number of iterations of scenario $s$, and any element $y^s_{t,c}$ represents the power value at temporal sample $t$ in basic channel $c$. From $Y^s$, we generate a binary matrix $X^s$ of same size through an occupancy indicator function, where any element $x^s_{t,c}$ represents whether channel $c$ was occupied at temporal sample $t$ (1) or not (0). Formally, $x^s_{t,c}=(y^s_{t,c}>\text{CCA}:1,0),\forall t,c$, where the clear channel assessment (CCA) is set to -83.5 dBm (or 150 10-bit RSSI units), corresponding to the common CCA threshold -82 dBm plus a safety margin of -1.5 dBm. While IEEE~802.11ac/11ax introduce different CCA levels for the primary and secondary channels, in this work, we consider a more restrictive approach by assuming the same threshold in order to fairly compare different channel bonding policies. The mean occupancy at a certain band $\mathcal{B}$ in scenario $s$ is simply defined as
	\begin{equation}
    	\bar{o}_\mathcal{B}^s = \frac{\sum_{t}\sum_{c\in \mathcal{B}}x_{t,c}^s}{N_\text{it}^s n_\text{s} |\mathcal{B}|}.
	\end{equation}
	
To provide meaningful experiments, we separately consider two 160-MHz bands composed of 8 basic channels: the U-NII-1\&2 and part of the U-NII-2c sub-bands, \mbox{$\mathcal{B}_\text{1\&2} = \{1,2,3,...,8\}$} and \mbox{$\mathcal{B}_\text{2c}= \{9,10,11,...,16\}$}, respectively. These sub-bands cover from channel 36 to 64 and from channel 100 to 128, respectively (see Figure~\ref{fig:warp_scheme}). Notice that these are the only sub-bands that allow to perform 160-MHz transmissions in the IEEE 802.11ac/ax channelization. Moreover, we focus on epochs (or periods) of duration  $T_\text{per} = 100$ ms for which the mean occupancy at such sub-bands is at least 5\%, i.e., $\bar{o}_\mathcal{B} \geq 0.05$, where $\mathcal{B} \in \{\mathcal{B}_\text{1\&2},\mathcal{B}_\text{2c}\}$.

\subsubsection{State Machine}
We develop a discrete state machine that characterizes how the channel-bonding BSS responds to each power sample (or temporal sample) $t$ according to the current state $S(t)$, and channel-bonding policy $\pi$, following the 802.11 standard. The set of possible states is $\mathcal{S} = \{\textit{Busy}, \textit{DIFS}, \textit{BO}, \textit{TX/RX}\}$. State \textit{Busy} indicates that the primary channel is busy, \textit{DIFS} represents the period before initiating the backoff process, the backoff counter is decreased during \textit{BO} state, and \textit{TX/RX} represents the actual frame transmission-reception (including the control frames RTS, CTS, and ACK, the DATA frame, and the SIFS periods in between). We represent the channel-bonding BSS $w$ as an AP and one or multiple clients that would perceive exactly the same spectrum activity as \sys captured in the measurement campaigns and must contend accordingly.


	\begin{figure}[ht]
		\centering
		\includegraphics[width=0.47\textwidth]{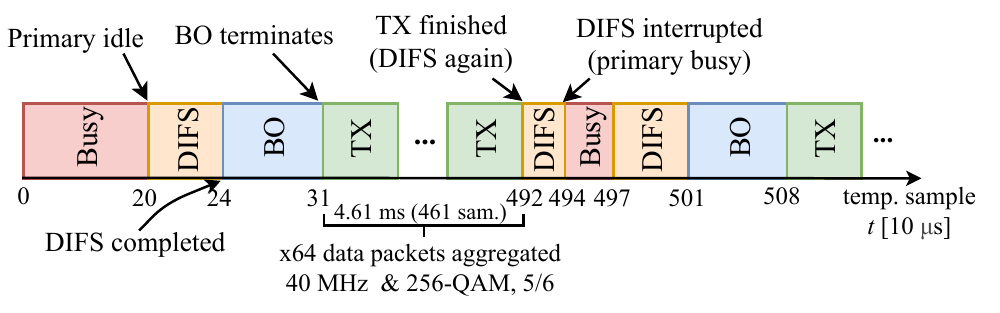}
		\caption{Example of the transitions between states.}
		\label{fig:states_example}
	\end{figure}

The set of basic channels selected for transmitting a frame depends both on the spectrum occupancy and on the selected channel-bonding policy $\pi$. Figure~\ref{fig:states_example} illustrates an example of the transitions between states. Empty slots have a duration $T_\text{slot} = \SI{10}{\micro\second}$ and we consider $T_\text{slot} = T_\text{s} = \SI{10}{\micro\second}$ rather than \SI{9}{\micro\second} (802.11's default value) to align the duration of an idle backoff slot with the sample duration. Hence, whenever the channel-bonding BSS  is in the backoff process at state \textit{BO}, every idle sample at the primary channel $p$ results in a backoff counter decrease of one empty slot. We use Wi-Fi parameters according to IEEE 802.11ax \cite{tgax2019draft} and assume 256-QAM modulation coding scheme regardless of the transmission bandwidth. The setup parameters are listed in Table~\ref{table:setup_simulations}.
After running the state-machine through all the temporal samples in the epoch, we compute the throughput $\Gamma$ as the number of data packets sent divided by the duration of the epoch $T_\text{per}$.

\begin{table}[h!]
	\small
	\centering
	\begin{tabular}{@{}ccc@{}}
		\toprule
		\textbf{Param.}  & \textbf{Description}        & \textbf{Value}                                                                                             \\ \midrule
		CCA    & CCA threshold  & -83.5 dBm  \\
		MCS		& MCS index  & 9 (256-QAM 5/6)  \\
		$b$    & \textit{basic} channel bandwidth  & 20 MHz  \\
		$L_\text{d}$    & Length of a data packet  & 12000 bit  \\
		$\mathcal{B}$    & Allocated set of \textit{basic} channels  & $\mathcal{B}\subseteq\mathcal{R}$ \\
		$p$    & Primary channel  & $p \in \mathcal{B}$ \\
		$N_\text{a}$    & Max. no. of agg. packets per frame  & 64 \\
		$T_\text{e}$                   & Duration of an empty slot                     & 10 \si\micro\si{s}      \\ 
		$T_\text{SIFS}$                   & SIFS duration                     & 20 \si\micro\si{s}      \\ 
		$T_\text{DIFS}$                   & DIFS duration                     & 30 \si\micro \si{s}      \\ 
		$T_\text{PIFS}$                   & PIFS duration                     & 30 \si\micro \si{s}      \\
		$T_\text{RTS}$                   & RTS duration                     & 50 \si\micro \si{s}      \\
		$T_\text{CTS}$                   & CTS duration                     & 40 \si\micro \si{s}      \\
		$T_\text{BACK}$                   & Block ACK duration                     & 50 \si\micro \si{s}      \\
		TXOP                   & Max. duration of a TXOP                     & 5 \si\milli \si{s}      \\
		CW$_\text{min}$                   & Min. contention window                    & 16       \\
		$m$                   &No. of backoff stages                    & 5       \\
		\bottomrule
	\end{tabular}
	\caption{Trace-driven setup.}
	\label{table:setup_simulations}
\end{table}
	
\subsubsection{Channel Bonding Policies and Response}
A channel-bonding policy $\pi$ selects the set of basic channels to aggregate at the end of the backoff provided that the primary channel is available. Namely, \emph{contiguous} channel bonding can select a set of channels both above and below the primary channel, provided they are consecutive. In contrast, \emph{non-contiguous channel bonding} can combine all available channels at the time the primary channel becomes available. We compare both policies in \S\ref{sec:cont_noncont}.

How will other BSS's respond to the channel bonding BSS? We consider that they will defer their transmissions. Namely, the channel bonding BSS needs the channels to be available only when its countdown timer expires. If the bonding BSS does transmit but the trace indicates that a channel  would have been occupied at some point during the transmission, we consider that such other BSS's will sense the bonding BSS and defer.

\subsection{Contiguous vs. Non-Contiguous} \label{sec:cont_noncont}
	
For each transmission, non-contiguous channel bonding can utilize additional channels as compared to contiguous, by ``skipping over'' the busy channels to find the next unused one. Here, we explore the gains of this flexibility as well as (in rare cases) the losses by comparing the throughput of contiguous and non-contiguous channel bonding in three load regimes: low ($\bar{o}_\mathcal{B} \leq 0.1$), medium ($0.1 < \bar{o}_\mathcal{B} \leq 0.2$) and high ($\bar{o}_\mathcal{B}>0.2$), respectively.
		
\begin{figure}[b!]
		\centering
		\includegraphics[width=0.4\textwidth]{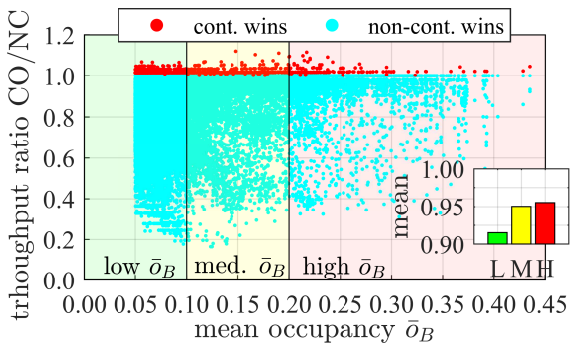}
		\caption{Throughput ratio of contiguous vs. non-contiguous channel bonding. The bar chart inset depicts the mean aggregated ratio for low (L), medium (M), and high (H) occupancy regimes.}
		\label{fig:results_ratio_cont_noncont}
\end{figure}
	
Figure~\ref{fig:results_ratio_cont_noncont} shows the the throughput ratio of contiguous to non-contiguous channel bonding $\Gamma_{\text{CO}}(p)/\Gamma_{\text{NC}}(p)$, where $\Gamma_\pi(p)$ is the throughput achieved by policy $\pi$ when selecting primary $p$ in a given period. 
We plot the ratio for all possible primaries in $\mathcal{B}_\text{1\&2}$ and $\mathcal{B}_\text{2c}$.

The data reveals two remarkable phenomenon. First,  contiguous outperforms non-contiguous in 2.5\% of the cases (albeit with a modest throughput difference of 1.9\%). But since non-contiguous is more flexible, how can it ever do worse? The answer is that the two policies result in different instants for transmission attempts. The contiguous policy occasionally (and quite randomly since the traces are the same) ends up with more favorable attempt instants. Nonetheless, in most cases, non-contiguous obtains higher throughput. For example, in many periods, at least one 20-MHz channel is idle during the whole period, which will always yield a gain for non-contiguous, but only sometimes yields a gain for contiguous bonding. In some cases, the difference can be quite high (e.g., a ratio of approximately 0.2 observed in low load). The origins of such extreme cases are the selection of the primary channel. Second, the bar chart inside Figure~~\ref{fig:results_ratio_cont_noncont} reveals that both contiguous and non-contiguous channel bonding perform quite close on average for all occupancy regimes (low, medium, and high), and especially for the latter, as high load results in far fewer bonding opportunities overall.	

\textit{Finding}: Non-contiguous almost always outperforms contiguous channel bonding and their throughput differences are occasionally over a factor of 5. Nonetheless, their average throughputs are quite similar, which may ultimately favor contiguous channel bonding, since it is simpler to implement.

\section{Conclusion}

    In this paper, we introduce \sys, an all-channel Wi-Fi spectrum analyzer for simultaneous measurement of all 24 20 MHz channels that allow channel bonding at 5 GHz. We present extensive measurement campaigns covering two continents, diverse areas, and many hours of signal strength  samples. The gathered dataset is a unique asset to find insights otherwise not possible to study. We leave the analysis of key factors like spectral correlation or bandwidth prevention as future work.

\section*{Acknowledgements}

We appreciate the generosity of Futbol Club Barcelona for letting us take measurements in the Camp Nou stadium. The authors would also like to thank Chia-Yi Yeh, Albert Bel, Álvaro López, and Marc Carrascosa for their contribution in taking the measurements. 
The work from S. Barrachina-Muñoz and B. Bellalta has been partially supported by CISCO (CG\#890107-SVCF), Maria de Maeztu Units of Excellence Programme (MDM-2015-0502), WINDMAL PGC2018-099959-B-I00 (MCIU/AEI/FEDER,UE), and SGR-2017-1188.
The research of E. Knightly was supported by Cisco, Intel, and by NSF grants CNS-1955075 and CNS-1824529.

\bibliographystyle{unsrt}
\bibliography{bib} 
	
\end{document}